\documentclass[final,3p,times,twocolumn]{elsarticle}

\usepackage{amsmath}
\usepackage{amssymb}

\biboptions{sort,compress,super}

\newcounter{bla}

\providecommand{\vect}[1]{\ensuremath{\boldsymbol{#1}}}
\providecommand{\pedex}[1]{\ensuremath{_\text{#1}}}
\providecommand{\apex}[1]{\ensuremath{^\text{#1}}}

\providecommand{\imag}{\ensuremath{\text{i}}}
\providecommand{\sgw}{\mbox{\textsc{SternheimerGW}}}

\usepackage[normalem]{ulem}

\journal{Computer Physics Communications}

\providecommand{\url}[1]{\texttt{#1}}
\usepackage{booktabs}
\usepackage{graphicx, xcolor}
\usepackage{sansmath}
\usepackage{dblfloatfix}

\begin{document}

\begin{frontmatter}



\title{\sgw: a program for calculating $GW$ quasiparticle band structures
and spectral functions without unoccupied states}


\author[]{Martin Schlipf\corref{author}}
\author[]{Henry Lambert\fnref{henry}}
\author[]{Nourdine Zibouche}
\author[]{Feliciano Giustino\corref{author}}

\cortext[author] {Corresponding authors.\\\textit{E-mail addresses:} martin.schlipf@gmail.com (M.~Schlipf), feliciano.giustino@materials.ox.ac.uk (F.~Giustino).}
\fntext[henry]{Present address: King’s College London, Physics Department, Strand, London WC2R 2LS, United Kingdom.}
\address{Department of Materials, University of Oxford, Parks Road, Oxford OX1 3PH, United Kingdom}

\begin{abstract}
The \sgw{} software uses time-dependent density-functional perturbation theory to evaluate 
$GW$ quasiparticle band structures and spectral functions for solids.
Both the Green's function $G$ and the screened Coulomb interaction $W$ are obtained
by solving linear Sternheimer equations, thus overcoming the need for a summation over
unoccupied states.
The code targets the calculation of accurate spectral properties by convoluting
$G$ and $W$ using a full frequency integration.
The linear response approach allows users to evaluate the spectral function at
arbitrary electron wavevectors, which is particularly useful for 
indirect band gap semiconductors and for simulations of angle-resolved photoelectron spectra.
The software is parallelized efficiently, integrates with version 6.3 of
Quantum Espresso, and is continuously monitored for stability using a test farm.
\end{abstract}

\begin{keyword}
first-principles calculations\sep many-body perturbation theory \sep solid state physics \sep linear response

\end{keyword}

\end{frontmatter}



\section*{Program Summary}

\begin{small}
\noindent
{\em Program Title:} \sgw                            \\
{\em Licensing provisions: GNU General Public License v3.0}   \\
{\em Programming language: Fortran 2003}                      \\
%
%
{\em Nature of problem:}
The lack of the exchange-correlation discontinuity in density-functional theory (DFT)
leads to a systematic underestimation of the band gap between conduction and
valence states.
Many-body perturbation theory in the $GW$ approximation provides an effective
solution to this problem, as well as other limitations faced by DFT in the description of electronic
excitations.
However, the $GW$ method comes with its own set of limitations:
(i) The perturbation of the system is typically expressed in terms of the 
unoccupied states, and achieving numerical convergence with respect to 
these states is often cumbersome.
Since the underlying DFT codes rely only on the occupied states, their
default behavior is often ill-suited to provide a sufficient amount of empty
states.
Combined with the large number of parameters that need to be converged, this
limits the use of $GW$ codes by non-expert users and automatic scripts.
(ii) Currently, $GW$ codes require that a homogeneous $\vect k$-point mesh is
used in the calculation.
Hence, features close to the band edges are often only accessible via
prohibitively expensive dense $\vect k$-point meshes or interpolation techniques.
(iii) To evaluate the frequency convolution of the Green's function $G$ 
and the screened Coulomb interaction $W$, many current $GW$ calculations rely on approximations
such as the plasmon-pole approximation, the analytic continuation, or the
contour deformation.
The relative merits and accuracy of the various approximations
are not fully understood.
\\
{\em Solution method:}
In \sgw{}, we address (i)
by replacing the summation over unoccupied
states with a linear response equation.
The solution depends on the occupied states only, and it employs linear response
algorithms already provided by the Quantum ESPRESSO suite to compute phonons
and related properties.
As an additional benefit, transforming the problem in a linear response equation
removes the restriction to homogeneous $\vect k$-point meshes (ii), so that the
$GW$ self-energy for any arbitrary point can be evaluated.
Finally (iii), we provide the possibility to perform a full frequency
convolution along the real frequency axis.
This feature can serve as a benchmark for approximate integrations using models or
analytic continuation.
\\
   \\
%
\end{small}

\section{Introduction}
\label{}

\begin{figure*}[b]
\centering
\includegraphics{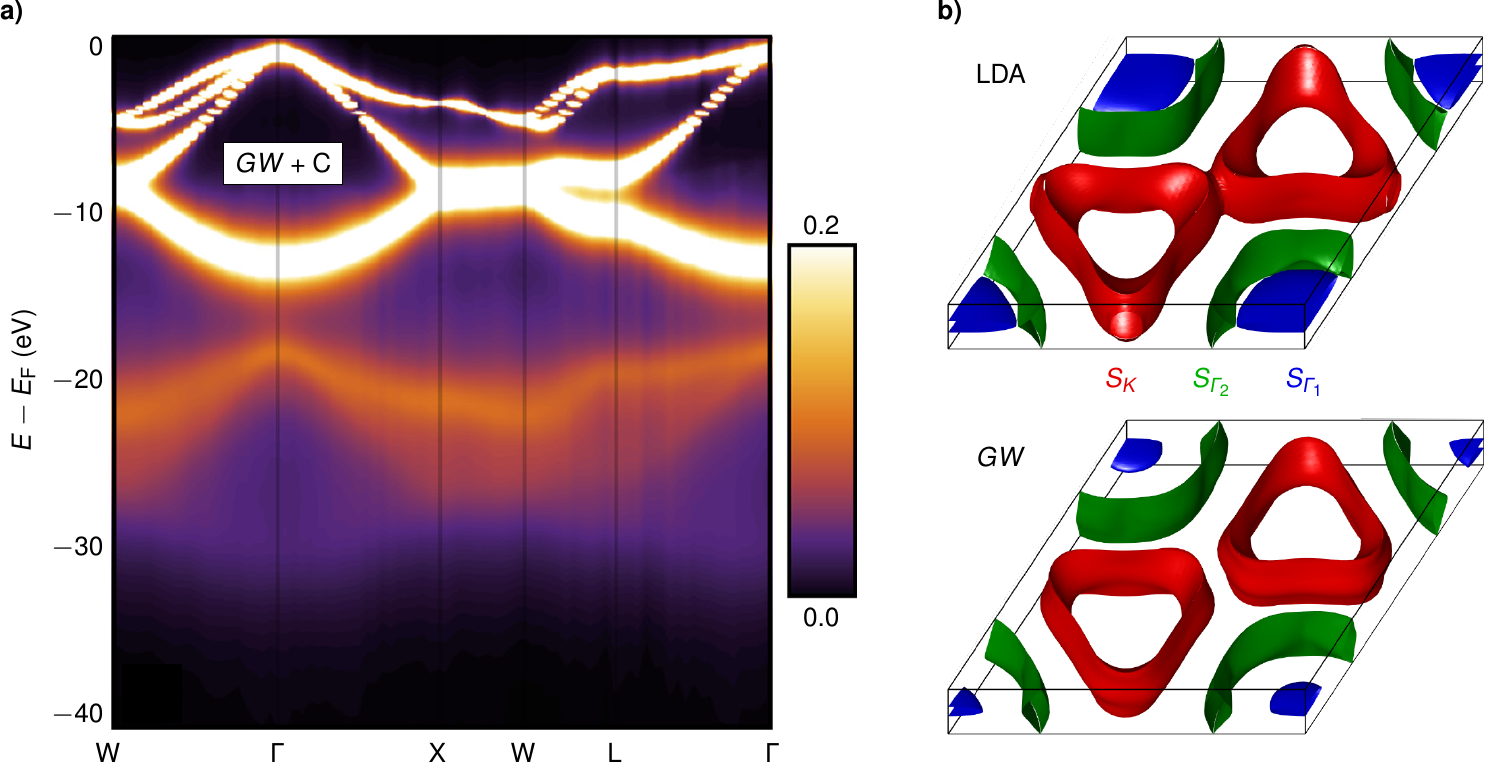}
\caption{Examples of recent calculations performed with the \sgw{} code.
{\bfseries a)} Spectral function of silicon (eV$^{-1}$) using the
$GW$ + cumulant expansion approach. Reproduced with permission from Ref.~\citenum{gumh16}.
{\bfseries b)} Comparison of the LDA and $GW$ Fermi surfaces of bulk NbS$_2$
near the $\Gamma$ (blue and green) and the K point (red). Reproduced with permission from 
Ref.~\citenum{Heil2018}.}
\label{fig:examples}
\end{figure*}

Kohn-Sham density function theory (DFT)\cite{hk64,ks65,Jones2015} is
a powerful and extremely popular formalism for studying the total energy of an 
interacting electron system in its ground state. When used in the study of
electronic excitations, such as the calculation of electron band structures 
and wavefunctions, DFT exhibits some
well known deficiencies, for example the lack of the exchange-correlation 
discontinuity in the exchange-correlation potential,\cite{pl83,ss83} which
leads to an underestimation of the band gap in insulators, and the excessive
delocalization of $d$ and $f$ electrons.
Applying many-body perturbation theory corrections in the $GW$ 
approximation\cite{hedin65,Hedin1969} allows one to correct some of these
deficiencies using the non-local and frequency-dependent electron self-energy $\Sigma(\omega)$.
Early numerical implementations of this method demonstrated an improved band gap
for diamond\cite{Strinati1980,Strinati1982} and similar improvement 
were subsequently shown for other typical semiconductors.\cite{hl86,Godby1986}
Since then, $GW$ has been used to accurately describe the band gaps for a variety of systems
ranging from solids to interfaces and molecules.\cite{ag98,Aulbur2000,orr02,Giantomassi2011}
It yields accurate band offsets\cite{Shaltaf2008}, defect energies,\cite{Rinke2009}
and improved effective masses.\cite{fvg15}
Recent developments focused on the self-consistency of the $GW$ method for
molecules\cite{Rostgaard2010,Caruso2012,Caruso2013} and solids,\cite{skf06} as well
as total energy calculations.\cite{Gatti2007,Beigi2010}

More recent developments of the $GW$ method relate to photoelectron spectroscopy.
Progress in high-energy resolution angle-resolved photoelectron spectroscopy
(ARPES) had led to a renewed interest in the spectral function of materials, in particular
the signatures of electron correlations and bosonic excitations.
These features were originally examined for the homogeneous electron gas,%
\cite{Lundqvist1967,Lundqvist1968,Hedin1969,Langreth1970,Baer1973}
but have recently been investigated both experimentally and theoretically
for charge carriers in semiconductors coupling to plasmons,%
\cite{Bostwick2010,Guzzo2011,Guzzo2012,Lischner2013,Lischner2014,Guzzo2014,Kas2014,
Lischner2015,Caruso2015,clg15, Caruso2016,Caruso2016a,gumh16,Caruso2018a} 
Fr\"ohlich polarons\cite{froe54,Chang2010,Moser2013,King2014,Story2014,
Antonius2015,Chen2015a,Wang2016,Cancellieri2016,Yukawa2016,Verdi2017,Nery2018}
or hybridizations of these excitations.\cite{Settnes2017,Riley2018}
While the $GW$ approximation does produce satellite features in the spectral
function, their strength is overestimated and their energy
is blue-shifted.%
\cite{Lundqvist1967,Lundqvist1968,Langreth1970,Kheifets2003,Guzzo2011}
This mismatch is intrinsic to the $GW$ approximation and cannot be overcome
by self-consistency.\cite{Holm1998,Kutepov2012,Caruso2013,Caruso2016}
Nevertheless, the $GW$ spectral function can provide a starting point for
the cumulant expansion, in which the satellite and quasiparticle spectral
function are convoluted and the position and weight of the satellite
features are corrected.%
\cite{Aryasetiawan1996,Holm1997,ag98,Hedin1999,Gumhalter2005,Gumhalter2012,Zhou2015,Verdi2017,Caruso2018}

At present, there are two major limitations to obtain accurate spectral functions
with the $GW$ approximation.
First, the frequency dependence of the screened Coulomb interaction is commonly
approximated by a Dirac delta function in the plasmon-pole model.\cite{hl86,Godby1989,Engel1993}
To increase the accuracy, more recent approaches employ an analytic 
continuation\cite{Daling1991,Rojas1995,Jin1999} or contour deformation.\cite{gss88,Kotani2002,laab03}
A detailed assessment of the accuracy of these techniques is still lacking
and may prove crucial for reliable calculations of quasiparticle 
lifetimes.\cite{Spataru2001,Park2009}
Secondly, most $GW$ implementations rely on a summation over the empty
Kohn-Sham states.\cite{hl86}
This requires a very accurate description of the unoccupied subspace and therefore
imposes additional constraints on the construction of pseudopotentials.
Furthermore, converging the eigenvalues occasionally demands the inclusion of a
considerable number of 
empty states\cite{Shih2010,Friedrich2011,Stankovski2011}, and the convergence
with respect to these states
is interdependent with other convergence parameters.\cite{Friedrich2011,Setten2017}
To address these issues, several techniques have been proposed%
\cite{Bruneval2008,Kang2010,Berger2010,Berger2012,Deslippe2013,Gao2016}
which improve the rate of convergence of the summations.
However, some of these methods introduce additional
convergence parameters.
Alternatively, one can obtain the dielectric matrix in 
density-functional perturbation theory\cite{bgt87,bgdg01} by solving
linear Sternheimer equations.\cite{rog97} 
This approach has been subsequently optimized by iterative diagonalization
to determine the most significant eigenvalues\cite{Wilson2008,Wilson2009,Pham2013}
or employing an optimal representation of the polarization.\cite{Umari2009,Umari2010}
One can extend this approach, so that both
the Green's function $G$ and the screened Coulomb interaction $W$
can be obtained by solving linear response equations.\cite{gcl10,Umari2009} 
This approach is referred to as the \emph{Sternheimer $GW$} method and has been
successfully applied to large systems with a single $\vect k$ point.\cite{Umari2009, Umari2010, Govoni2015}

In this work, we develop an implementation for solids building on the methodology
described in Ref.~\citenum{gcl10} and~\citenum{lg13}.
Our implementation, called \sgw{}, features the plasmon pole model, the analytic
continuation, as well as the real frequency integration on the same footing, and is 
therefore suitable to assess the accuracy of these techniques.
The frequency dependent self-energy allows us to access spectral properties from 
first principles.\cite{Park2009}
Two illustrative examples of the application of \sgw{} are shown in Fig.~\ref{fig:examples}.
In Fig.~\ref{fig:examples}a we show the complete wavevector-dependent spectral function of silicon.
For these results the frequency-dependent self-energy obtained with \sgw{} is
used as a starting point for a cumulant expansion, to obtain both quasiparticle
band structure and plasmon replicas with the correct energy and intensity.\cite{clg15, gumh16}
In Fig.~\ref{fig:examples}b, we show the use of \sgw{} to evaluate the Fermi surface 
of the transition metal dichalcogenide NbS$_2$.\cite{Heil2018}
In this system, the change of density of states at the Fermi energy
due to many-body perturbation theory corrections is important to obtain
more accurate predictions for the superconducting transition
temperature.\cite{heil17}

\section{Description of the code}

\subsection{Kohn-Sham reference system}

To evaluate the many-body perturbation correction, we start from an unperturbed reference
system within Kohn-Sham density functional theory.
The important ingredients of this reference system for our calculation are
the exchange-correlation potential $V\apex{xc}$ and the Kohn-Sham Hamiltonian
$\hat H_{\vect k}$ along with its eigenvalues $\varepsilon_{n\vect k}$
 and wavefunctions $\phi_{n\vect k}(\vect r)$.
Here $n$ is a band index, $\vect r$ is the position vector, 
and $\vect k$ a Bloch vector in the Brillouin zone
of the crystal.
We employ a planewaves basis set, and express the wavefunction
$\phi_{n\vect k}(\vect r)$ in terms of
the expansion coefficients $u_{n\vect k,\vect G} $:
\begin{equation}
  \phi_{n\vect k}(\vect r) = \frac{1}{\sqrt{\Omega}}\sum_{\vect G} u_{n\vect k,\vect G} \exp\bigl[\imag (\vect k + \vect G) \cdot \vect r\bigr]~,
\end{equation}
where $\vect G$ denotes reciprocal lattice vectors and $\Omega$ is the
volume of the unit cell.
In a planewaves basis only the kinetic energy part of the Hamiltonian 
and the non-local part of the pseudopotentials are represented in reciprocal space;
the local Kohn-Sham potential is applied in real space via a fast Fourier transform.

\subsection{Green's function}

From the Kohn-Sham Hamiltonian, we obtain the electronic Green's function,
which is one of the ingredients to evaluate the $GW$ self-energy. The Green's
function is given by the linear equation:
\begin{equation} \label{eq:green}
  \forall_{\vect k,\vect G'} \qquad \mathcal{L}\apex{G}_{\vect k}(\omega)~G_{\vect G\vect G'}(\omega)
  = -\delta_{\vect G\vect G'}~.
\end{equation}
Here, $\omega$ is a complex frequency, chosen to satisfy the time ordering,
$\mathcal{L}\apex{G}$ is the linear operator for the Green's function:
\begin{equation} \label{linop:green}
  \mathcal{L}\apex{G}_{\vect k}(\omega) = \hat H_{\vect k} - \hbar \omega~,
\end{equation}
 and $\delta_{\vect G\vect G'}$ is the Kronecker $\delta$.
Most $GW$ codes expand Eq.~(\ref{eq:green}) in the basis of the eigenstates of the
Hamiltonian and solve it by summing over many unoccupied states.
In contrast, in \sgw{} we utilize an iterative Krylov subspace solver\cite{from03}
to explicitly find the solution to the linear equation~\eqref{eq:green}.
We employ the fact that shifted linear problems described by the linear 
operator $\mathcal{L}\apex{G}$ differing only in the frequency $\omega$
span the same Krylov subspace. This property allows us to solve the linear problem
for all frequencies at the computational complexity of a single frequency.

\subsection{Dielectric matrix}

The dielectric matrix of a system can be evaluated from the electron
density response to a planewave perturbation:
\begin{equation}
  \epsilon_{\vect q, \vect G\vect G'}(\omega) = \delta_{\vect G\vect G'} 
  - \Delta n_{\vect q, \vect G\vect G'}(\omega)~,
\end{equation}
where $\omega$ is the frequency.
The calculation of the density response is performed by starting
from the linear variation of the wavefunctions in real space, and then
carrying out a Fourier transform into reciprocal space:
\begin{equation}
  \Delta n_{\vect q,\vect r\vect G'}(\omega) = 2 \frac{1}{N_{\vect k}}
  \sum_{n\vect k,\pm}\apex{occ.} u_{n\vect k,\vect r}^\ast
  \Delta u_{n\vect k + \vect q,\vect r\vect G'}(\pm\omega)~,
\end{equation}
where the factor 2 accounts for the spin degeneracy and we are assuming a
uniform grid of $N_{\vect k}$ $\vect k$-vectors in the Brillouin zone.
The change of the wavefunction coefficients is obtained from the Sternheimer equation,
which can be written as:
\begin{equation} \label{eq:wavef}
  \forall_{n\vect k,\vect G} \quad \mathcal{L}\apex{C}_{\vect k + \vect q}
  \Bigl(\frac{\varepsilon_{n\vect k + \vect q}}{\hbar} + \omega\Bigr)
  \Delta u_{n\vect k + \vect q,\vect G\vect G'}(\omega) = \delta u_{n\vect k,\vect G\vect G'}~.
\end{equation}
Since the linear operator of the Coulomb response,
\begin{equation}
  \mathcal{L}\apex{C}_{\vect k}(\omega) = \hat H_{\vect k} - \hbar \omega + \hat P\apex{occ},
\end{equation}
is very similar to the one for the Green's function \eqref{linop:green}, the
same Krylov subspace solvers can be employed.
The only difference with respect to \eqref{linop:green} is that now
we have an additional projector on the occupied states $\hat P\apex{occ}$; this
term improves the condition number of the linear operator.
The right-hand side of \eqref{eq:wavef} is given by a Fourier transform
of:
\begin{equation}
  \delta u_{n\vect k,\vect G\vect r'} = \bigl(1 - \hat P\apex{occ}\bigr) 
  e^{i \vect G \cdot \vect r'} u_{n\vect k,\vect r'},
\end{equation}
and describes a driving field corresponding to a periodic perturbation.

\subsection{Correlation self-energy} \label{sec:corr}

For the correlation self-energy, we combine the inverse of the dielectric
matrix with the truncated Coulomb potential (see Sec.~\ref{sec:trunc}):
\begin{equation}
  \bar W_{\vect q, \vect G \vect G'}(\omega) = 
  v_{\vect q + \vect G}~\bigl(\epsilon_{\vect q, \vect G \vect G'}^{-1}(\omega) - \delta_{\vect G \vect G'}\bigr)~,
\end{equation}
where $\bar W$ is the correlation part of the screened Coulomb interaction.
To overcome numerical issues associated with the inversion of the dielectric
matrix for $q \approx 0$,\cite{hl87i} we {evaluate that element for a
small but finite $\vect q$\cite{gamma-correct} and} set the off-diagonal elements to zero
if $\vect q + \vect G = \vect 0$ or $\vect q + \vect G' = \vect 0$.\cite{Martin-Samos2009}
Subsequently,
we perform a Fourier transform of $G$ and $\bar W$ to real space, because the
self-energy can be expressed as a product there:
\begin{equation} \label{eq:corr}
  \Sigma\apex{c}_{\vect k,\vect r \vect r'}(\omega) = 
  \sum_{\omega'\vect q} \frac{\alpha w_{\omega'} w_{\vect q}}{2 \pi}
  G_{\vect k - \vect q,\vect r  \vect r'}(\omega') \bar W_{\vect q,\vect r \vect r'}(\omega' - \omega)~.
\end{equation}
$w_{\vect q}$ and $w_{\omega'}$ are weights for $\vect q$ and $\omega$ integration
respectively.
The factor $\alpha$ depends on whether the frequency integration is performed
along the real axis ($\alpha = \imag$) or along the imaginary axis ($\alpha = -1$).
We transform the resulting self-energy back to reciprocal space to evaluate
the expectation values with the Kohn-Sham wavefunctions.

\subsection{Exchange self-energy}

To evaluate the exchange self-energy, we evaluate the Fock potential by summing 
over all occupied wavefunctions:
\begin{equation}
  \Sigma\apex{x}_{\vect k, \vect G\vect G'} = -
  \sum_{n\vect q,\vect G''}\apex{occ.} \frac{w_{n\vect k - \vect q}}{\Omega} u^\ast_{n\vect k-\vect q,\vect G' - \vect G''}
  u_{n\vect k-\vect q,\vect G-\vect G''} v_{\vect q + \vect G''},
\end{equation}
where the weight $w_{n\vect k - \vect q}$ contains the weight of the $\vect q$ integration.
The Coulomb potential is truncated as described in the following section.

\begin{figure*}[b]
\centering
\includegraphics{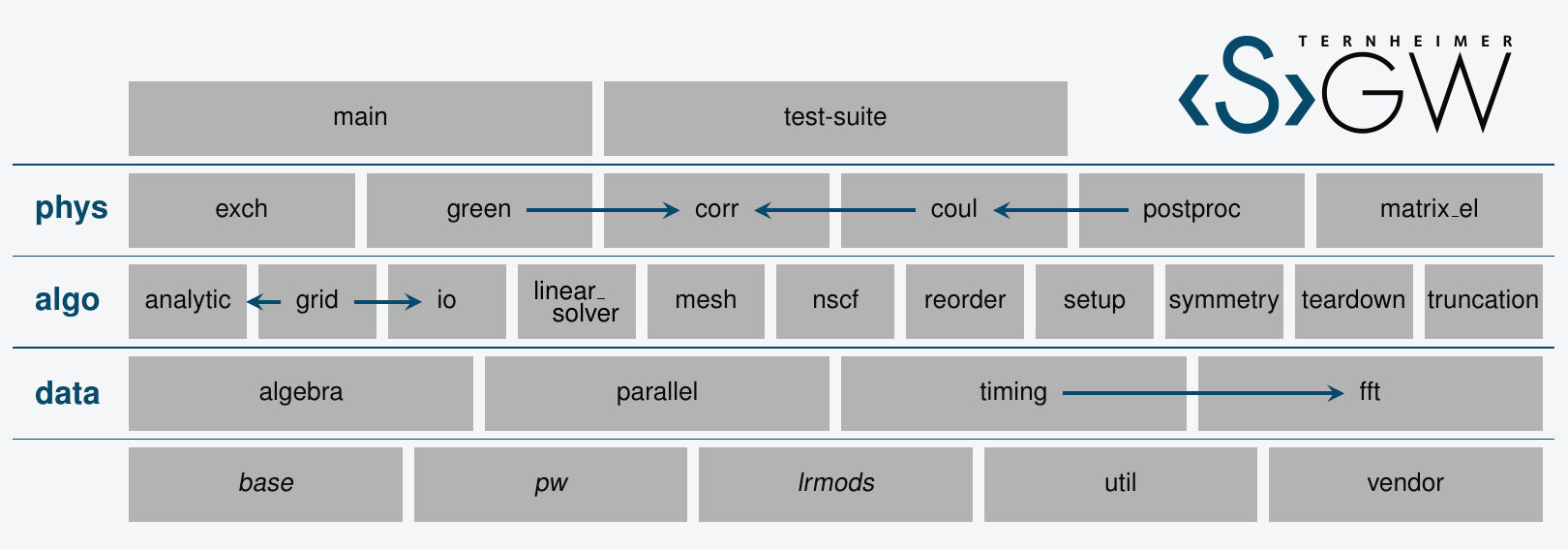}
  \caption{Multitier architecture of the \sgw{} code. The lowest tier includes external
  libraries as well as a utility package. The packages written in italic are not
  distributed with the \sgw{} code and require the Quantum ESPRESSO software.
  In the \emph{data} tier, we implement wrappers to these
  libraries. We combine these packages in the \emph{algo} tier to provide the necessary
  data structures and algorithms, which are used in the \emph{phys} tier to solve the physical
  problem. The highest tier interacts with the user and is used for testing the program.
  Arrows indicate dependencies within the same tier. 
  }
  \label{fig:arch}
\end{figure*}

\subsection{Truncation} \label{sec:trunc}
Owing to the long-range nature of the Coulomb potential, the Fourier transform of 
this quantity diverges for small
arguments $q \rightarrow 0$, leading to instabilities in the convergence
with respect to the grid in the Brillouin zone. Furthermore, when describing
systems of reduced dimensionality, such as slabs or nanowires, it is advantageous to
truncate the Coulomb potential along the non-periodic direction in order to avoid spurious
interactions between periodic images. In order to address these issues, it is common
practice to truncate\cite{beigi06,Rozzi2006,sa13}
the Coulomb potential at a certain
distance. This approach yields a Fourier transform with
a finite value in the limit $q \rightarrow 0$.
In \sgw, we implement three different truncation schemes:
First, one can truncate the Coulomb potential at a distance $R\pedex{cut}$
(\emph{spherical truncation}) resulting in a potential:
\begin{equation}
  v_{\vect q} = \frac{e^2}{4\pi\epsilon_0} \frac{4\pi}{q^2} \bigl[1 - \cos(q R\pedex{cut})\bigr]~.
\end{equation}
Secondly, one can truncate the Coulomb interaction at a height $z\pedex{cut}$ 
(\emph{slab truncation}), which yields the following potential:\cite{beigi06}
\begin{equation}
  v_{\vect q} = \frac{e^2}{4\pi\epsilon_0} \frac{4\pi}{q^2}
  \Bigl[1 - \exp(\sqrt{q_x^2 + q_y^2} z\pedex{cut})\Bigr] \cos(q_z z\pedex{cut})~.
\end{equation}
Lastly, one can truncate the Coulomb interaction in the Wigner-Seitz supercell.\cite{sa13}
In this latter case, since an analytic expression for the potential is not known, 
we tabulate the results of the
truncation and look up the values for the relevant vectors as necessary.

\subsection{Analytic continuation}
Quantities such as the dielectric matrix and the self-energy exhibit significant
structure along the real axis, for example due to electron-hole excitations.
In order to improve numerical stability and accuracy, it can be
advantageous to evaluate quantities along the
imaginary axis and perform an analytic continuation to the real axis.
In \sgw, there are two levels where such an analytic continuation
may be employed.
On the one hand, one can evaluate the dielectric function on the imaginary axis,
perform an analytic continuation to the real axis, and subsequently convolute it
with the Green's function according to Eq.~\eqref{eq:corr}.
On the other hand, one can resolve the convolution along the imaginary
axis, thereby obtaining a self-energy at imaginary frequency values.
The self-energy at real frequencies is then obtained via analytic continuation.
To perform the analytic continuation, one can use a Pad\'{e} expansion\cite{vs77}
or the adaptive Antoulas-Anderson (AAA) algorithm.\cite{Nakatsukasa2018} 
In general, determining the Pad\'{e} approximant using a continued fraction expansion
suffers from numerical instabilities, since a very high precision
(number of significant digits) is required. In contrast, the AAA algorithm
relies on a singular-value decomposition, and it refines iteratively
the data points used to construct the approximant, until the largest 
deviation falls below a specified threshold.

\subsection{Spectral function}

The spectral function is related to the imaginary part of the retarded Green's function:
\begin{equation}
  A_{\vect k,\vect G\vect G'}(\omega) = -\frac{1}{\pi} \text{Im}\bigl(G_{\vect k,\vect G\vect G'}(\omega)\bigr),
\end{equation}
and represents a $\vect k$-resolved many-body density of states.
Since the Green's function is the resolvent of the many-body Hamiltonian, this leads to:
\begin{equation}
  A_{\vect k, \vect G \vect G'}(\omega) = -\frac{1}{\pi} \text{Im} \Bigl(\hbar \omega - H_{\vect k} - \Sigma\apex{xc}_{\vect k}(\omega) + V_{\vect k}\apex{xc} \Bigr)^{-1}_{\vect G\vect G'}~.
\end{equation}
Approximating the $GW$ wavefunctions by their Kohn-Sham counterpart $\phi_{n\vect k}(\vect r)$,
the diagonal part of the spectral function in the Kohn-Sham basis can be expressed as:
\begin{equation}
  A_{n\vect k}(\omega) = \frac{-\text{Im}{\Sigma\apex{c}_{n\vect k}(\omega)}}{\pi \Bigl\lvert\hbar \omega - \varepsilon_{n\vect k} - \Sigma\apex{xc}_{n\vect k}(\omega) + V_{n\vect k}\apex{xc} \Bigr\rvert^2}~.
\end{equation}
The peaks of the spectral function correspond to the quasiparticle eigenvalues.
Linearizing the equation in the vicinity of the eigenvalue $\varepsilon_{n\vect k}$
results in the following quasiparticle eigenvalue:
\begin{equation}
  \varepsilon_{n\vect k}\apex{QP} = \varepsilon_{nk} + Z_{n\vect k} 
  \Bigl(\text{Re}\Sigma\apex{xc}_{n\vect k}(\varepsilon_{nk} / \hbar) - V_{n\vect k}\apex{xc}\Bigr)~,
\end{equation}
where $Z_{n\vect k} = 1 - \hbar^{-1} \text{Re} \bigl(\partial \Sigma_{n\vect k}\apex{c} / \partial \omega\bigr)$.

\begin{figure*}[b]
\centering
\includegraphics{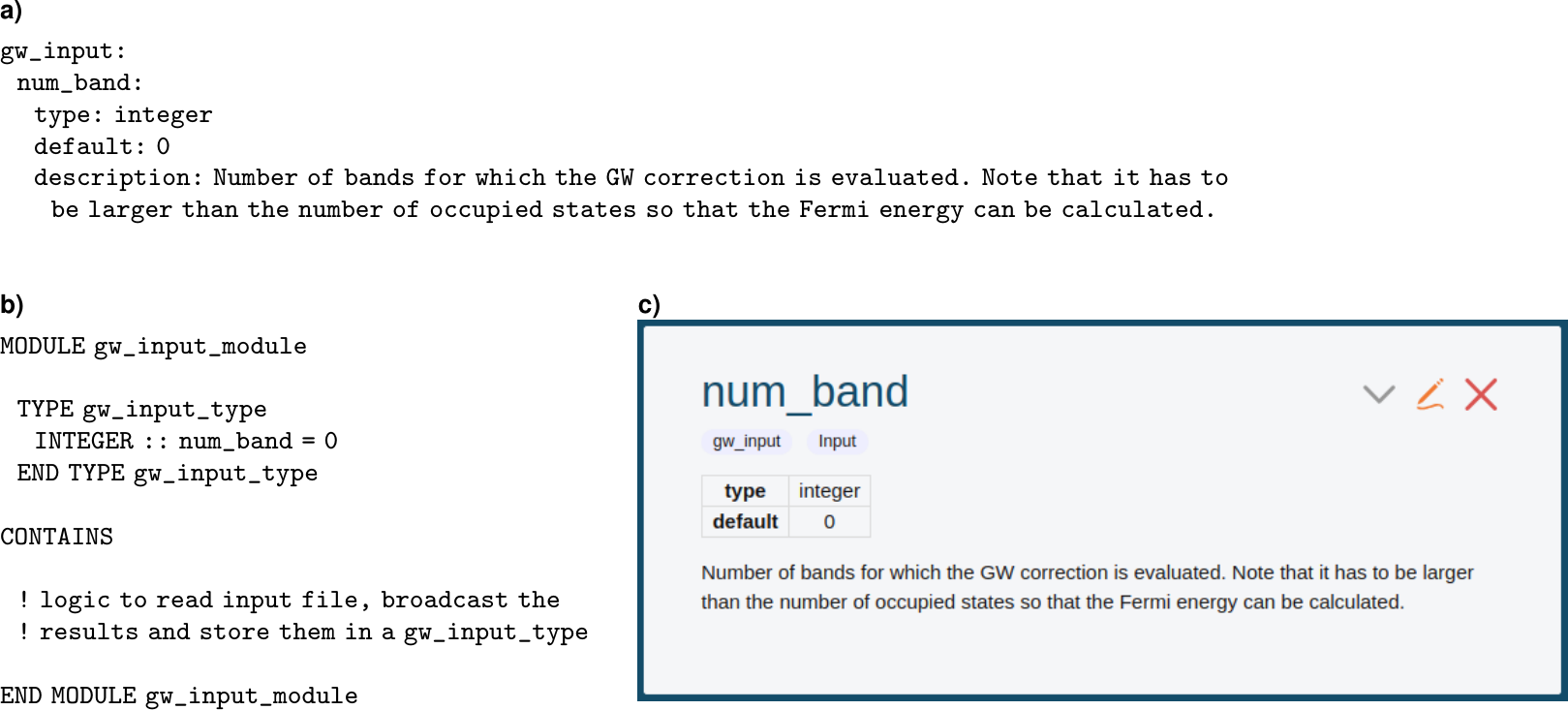}
\caption{Automatic generation of the user-input logic in SternheimerGW: 
{\bfseries a)} The gw\_input.yml
file defines which input variables are used in \sgw{} along with their type,
default value, and a description. {\bfseries b)} A script processes the YAML file to generate
a Fortran file containing a type with all input variables and the logic to read them.
{\bfseries c)} The YAML file is also processed to generate the description of the input
variables on the website \url{http://sternheimergw.org}.}
\label{fig:user}
\end{figure*}

\subsection{Software design}

\paragraph{Architecture} 
In Fig.~\ref{fig:arch}, we illustrate the multitier architecture of \sgw{}.
The code builds on functionality provided by the Quantum Espresso software package\cite{QE2009,QE2017}
and provides additional functionality handling low-level operations in the \emph{data} tier.
On top of this layer, we have an \emph{algo} tier handling 
the interface with Quantum Espresso.
The \emph{algo} tier also
provides integration grids, linear solvers, Coulomb truncation, and analytic
continuation functionalities.
In the \emph{phys} tier, we evaluate the $GW$ self-energy and matrix elements,
employing the functionality of the lower lying tiers.
The highest tier provides the user interface and the infrastructure for
testing the software.

\paragraph{User input}
The typical approach for specifying the user interface of a code requires that changes
to input variables are manually translated into an update of the documentation.
As this scheme poses the risk that the documentation becomes outdated, we employ
an inverse scheme in \sgw.
The core file controlling the user input is a human- and machine-readable YAML file 
documenting all input variables (see example in Fig.~\ref{fig:user}a).
Starting from the documentation, we run a script to generate the Fortran file
that processes the input file (Fig.~\ref{fig:user}b).
As a consequence, introducing a new input variable requires an update of
the documentation. This strategy allows us to make sure that the code and
the documentation are always in sync.
Furthermore, we process the YAML file to generate the online documentation of
\sgw{} (Fig.~\ref{fig:user}c) on the website \url{http://sternheimergw.org}.

\paragraph{Continuous integration and testing}
\begin{table}
\centering
\caption{Combined code coverage of integration and unit tests, i.e.,
the fraction of F90 lines and subroutines of the code that are executed at
least once by the test set. The higher tiers are covered exclusively by
the integration tests; for the lower tiers unit tests play a large role.}
\label{tab:codecov}
\begin{sansmath}\sffamily\footnotesize%
\begin{tabular}{l *{4}{r}}
  \toprule
  & \multicolumn{1}{c}{main} & \multicolumn{1}{c}{phys} & \multicolumn{1}{c}{algo} & \multicolumn{1}{c}{data} \\
  \midrule
  \emph{F90 lines} \\
  integration tests & 100.0\% & 92.1\% & 70.3\% & 31.4\% \\
  unit tests        &   0.0\% &  0.0\% & 41.3\% & 74.3\% \\
  combined          & 100.0\% & 92.1\% & 86.7\% & 92.3\% \\
  \midrule
  \emph{subroutines} \\
  integration tests & 100.0\% & 97.3\% & 64.6\% & 30.0\% \\
  unit tests        &   0.0\% &  0.0\% & 53.3\% & 80.0\% \\
  combined          & 100.0\% & 97.3\% & 87.7\% & 93.3\% \\
  \bottomrule
\end{tabular}
\end{sansmath}
\end{table}
To ensure the long-term stability of the code and its alignment with the
Quantum Espresso suite, we employ two techniques.
First, we examine whether the code compiles and reproduces a set of
reference results using a buildbot test farm.\cite{testcode, testfarm}
These integration tests involve 12 different calculations to check most of the
functionality of \sgw{}.
Secondly, we perform unit testing of some individual packages of the code, by
benchmarking the parallelization, the algebra, the analytic continuation,
the linear solver, and the input/output (io) package independently from the rest
of the code.
For the unit testing, at variance with
the tests of the complete code, we compare our implementation 
with analytic results.
The unit tests are implemented in the pFUnit framework.\cite{Rilee2014}
Overall, all these tests cover at least 80\% of the code in each tier
(see Table~\ref{tab:codecov}).

\subsection{Parallelization}

\begin{figure}
\includegraphics{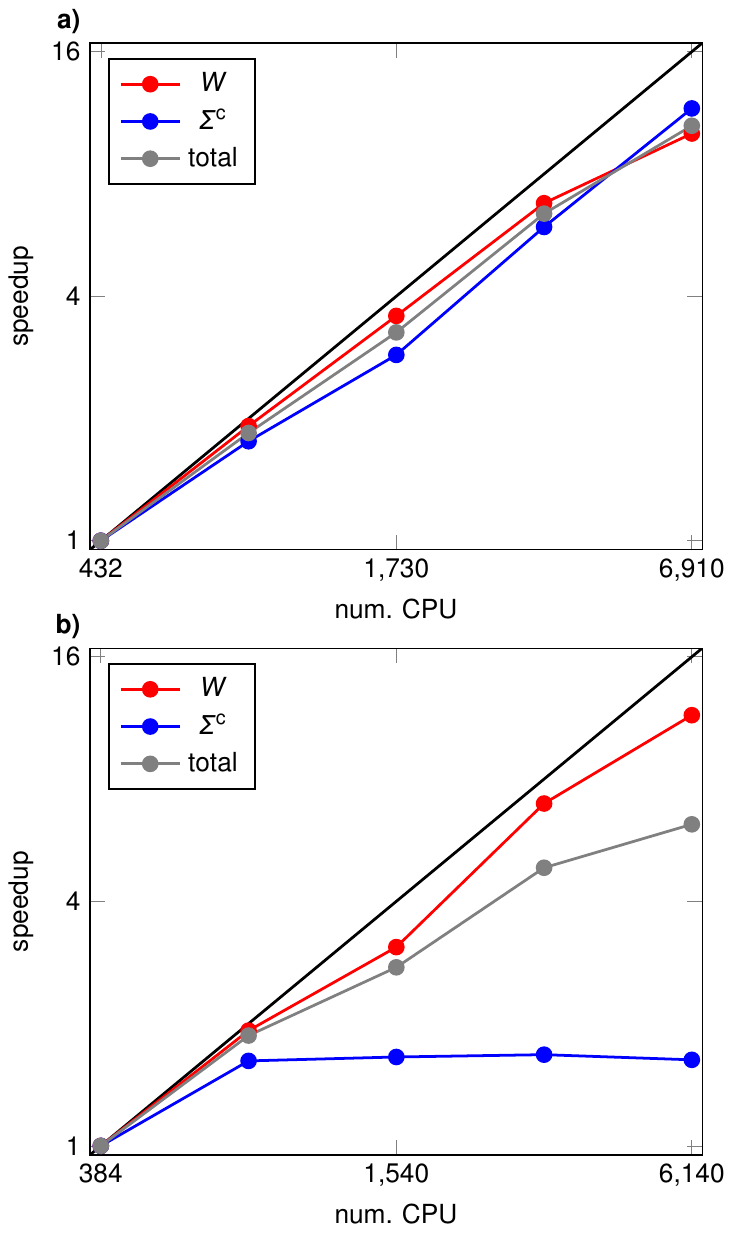}
\caption{Parallel efficiency of the \sgw{} code for the screened Coulomb 
interaction $W$ (red), the correlation self-energy $\Sigma\apex{c}$ (blue),
and the whole code (gray) using a) pool parallelization and b) image parallelization.
Measured for a bulk calculation of MgO on MareNostrum 4, BSC-CNS (Spain)
with converged parameters as listed in Table~\ref{tab:convergence}.}
\label{fig:para}
\end{figure}

$GW$ calculations are computationally much more demanding than Kohn-Sham DFT
calculations.
To allow for a reasonable time to solution, we employ two different parallelization
levels in \sgw. As we inherit the distribution logic from Quantum Espresso,
we refer to these levels as \emph{pool} and \emph{image} parallelization.
In Fig.~\ref{fig:para}, we compare the performance of these parallelization
levels for a bulk system of MgO for different parts of the code.
With the pool parallelization, we distribute different $\vect k$ or $\vect q$ points
to different processors.
This method is very efficient for both $G$ and $W$, if the number of these points 
is a multiple of the number of CPUs.
The image parallelization distributes $\vect G$ vectors used in the linear response
and Fourier transform to different CPU.
For the calculation of $W$, this parallelization is very efficient because 
the solution of the linear-response equation for each $\vect G$ vector
is independent of the other $\vect G$ vectors.
When evaluating $\Sigma\apex{c}$ the speedup of this parallelization saturates when
the number of grid points along the $z$ axis reduces to one per CPU.
The benefit of the image parallelization is that it reduces the memory footprint
of the code.

\section{Examples}

\subsection{Benchmark results}

\begin{table}
  \centering
  \setlength{\tabcolsep}{4.5pt}
  \caption{Comparison of the band gap of a selected set of materials with
  reference results from the literature. All values are given in eV.
  Numerical parameters used for the calculations are listed in Table~\ref{tab:convergence}.}
  \label{tab:benchmark}
  \begin{sansmath}\sffamily\footnotesize%
  \begin{tabular}{l *{5}{c}}
    \toprule
     & Present & Ref.~\citenum{sk07} & Ref.~\citenum{fbsbs12} & Ref.~\citenum{Chen2014} & Ref.~\citenum{Chen2015b} \\
    \midrule
    BN   & 6.27 & 6.10 & 6.20 &      & 6.19 \\
    C    & 5.53 & 5.50 & 5.62 & 5.62 & 5.59 \\
    CdS  & 2.20 & 2.06 & 2.18 &      & 2.31 \\
    GaAs & 1.23 & 1.30 & 1.31 & 1.21 & 1.10 \\
    Ge   & 0.78 &      & 0.75 & 0.63 & 0.50 \\
    MgO  & 7.13 & 7.25 & 7.17 & 6.71 & 7.08 \\
    Si   & 1.17 & 1.12 & 1.11 & 1.26 & 1.17 \\
    \bottomrule
  \end{tabular}
  \end{sansmath}
\end{table}

\begin{table*}\centering
\caption{Convergence parameters used in the calculation of the $GW$ corrections
reported in Table \ref{tab:benchmark}: The pseudopotentials (PP) are from the
SG15 library\cite{sg15} or the stringent ones from \url{pseudo-dojo.org};\cite{Setten2018} the
energy cutoffs are employed for the DFT self-consistent field cycle (scf), the
exchange (x) or the correlation (c) self energy; the integration grids for
the self-energy and the density response are homogeneous grids with $N_q$ and
$N_k$ points; the coarse frequency mesh for $W$ are obtained from a $(N_\omega - 1)$-node
Gauss-Laguerre quadrature along the imaginary axis including the origin,
and extending up to $\hbar\hat\omega$; and
the dense mesh for $W$ as well as the coarse one for $\Sigma$ are automatically
constructed by the code. The row with the dagger contains the loose values used
for the convergence test
and the row with the asterisk is used for the convolution
along the real frequency axis.}
\label{tab:convergence}
\begin{sansmath}\sffamily\footnotesize%
\begin{tabular}{*{15}{c}}
\toprule
&& \multicolumn{3}{c}{energy cutoffs (Ry)} &&& \multicolumn{2}{c}{coarse, $W$}
& \multicolumn{2}{c}{dense, $W$} & \multicolumn{2}{c}{coarse, $\Sigma$} &
\multicolumn{2}{c}{threshold} \\
material & PP & $E\pedex{cut}\apex{scf}$ & $E\pedex{cut}\apex{x}$ & $E\pedex{cut}\apex{c}$ & 
$N_q^{1/3}$ & $N_k^{1/3}$ & $N_{\omega}$ & $\hbar\hat\omega$ (eV) &
$N_{\omega}$ & $\hbar\hat\omega$ (eV) & $N_{\omega}$ & $\hbar\hat\omega$ (eV) &
$t_W$ & $t_G$ \\
\midrule
BN   & SG15 & 60 & 50 & 40 & 8 & 8 & 31 & 104 &  80 & 320 & 20 & 150 & $10^{-4}$ & $10^{-5}$ \\
C    & SG15 & 70 & 50 & 24 & 8 & 8 & 31 & 104 &  50 & 280 & 30 & 180 & $10^{-4}$ & $10^{-5}$ \\
CdS  & Dojo & 60 & 55 & 25 & 8 & 8 & 31 & 104 & 150 & 150 & 30 & 180 & $10^{-4}$ & $10^{-5}$ \\
GaAs & Dojo & 90 & 70 & 40 & 8 & 8 & 31 & 104 & 130 & 130 & 20 & 150 & $10^{-4}$ & $10^{-5}$ \\
Ge   & Dojo & 90 & 70 & 25 & 8 & 8 & 31 & 104 & 130 & 130 & 20 & 150 & $10^{-4}$ & $10^{-5}$ \\
MgO  & SG15 & 70 & 65 & 30 & 6 & 8 & 31 & 104 &  50 & 280 & 30 & 180 & $10^{-4}$ & $10^{-5}$ \\
Si   & SG15 & 20 & 19 & 15 & 8 & 8 & 31 & 104 &  50 & 280 & 40 & 200 & $10^{-4}$ & $10^{-6}$ \\
\midrule
BN$^\dagger$ & SG15 & 60 & 30 & 10 & 1 & 4 & 21 & 67 & 50 & 50 & 2 & 100 & $10^{-4}$ & $10^{-5}$ \\
Si$^\ast$ & SG15 & 16 & 15 & 12 & 8 & 8 & 201 & 768 & 800 & 100 & 351 & 35 & $10^{-10}$ & $10^{-5}$ \\
\bottomrule
\end{tabular}
\end{sansmath}
\end{table*}

\begin{figure*}
\centering
\includegraphics{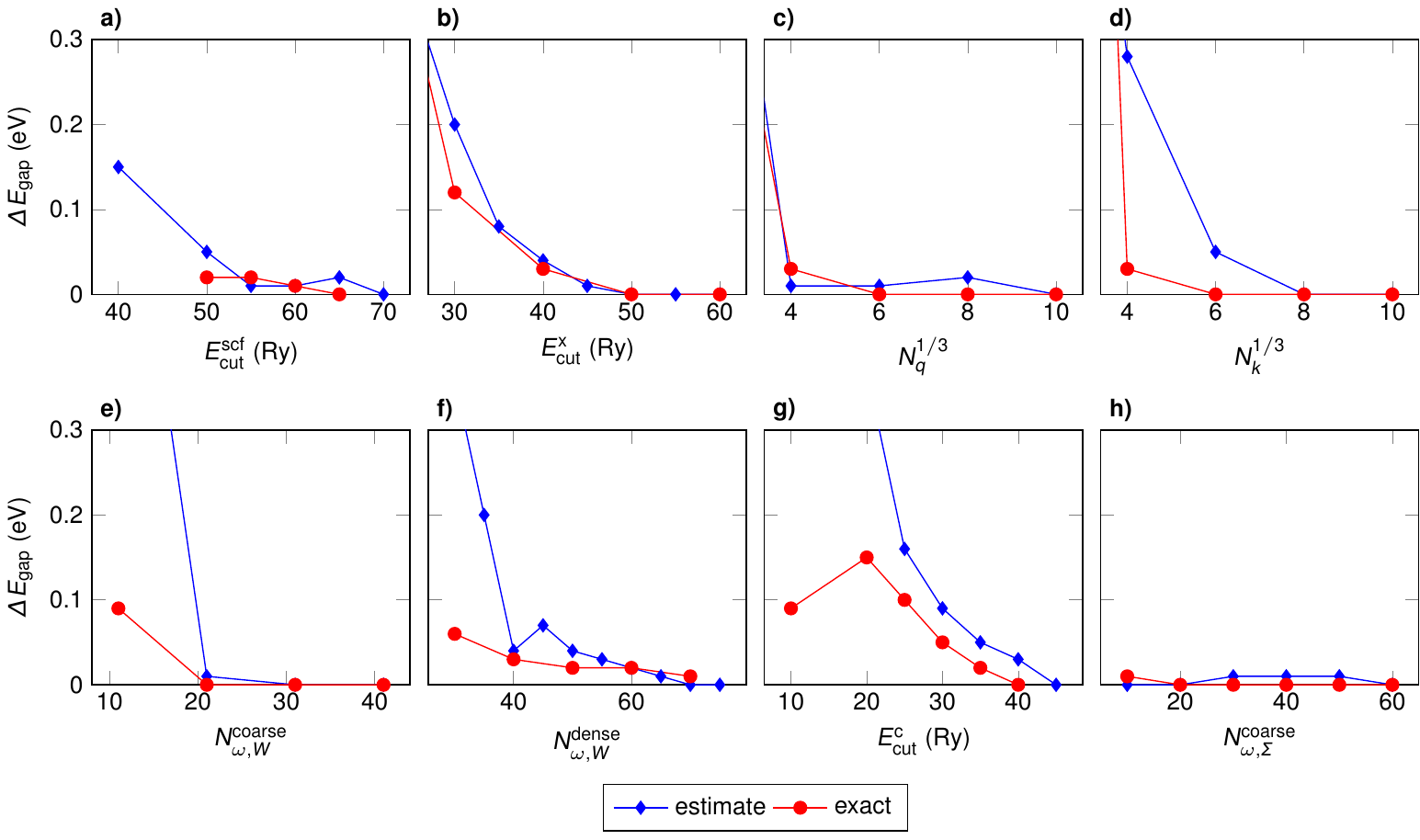}
\caption{Convergence of the band gap of BN in the zincblende 
structure with respect to various convergence parameters in \sgw.
The quantity $\Delta E\pedex{gap}$ indicates the difference between the band gap with a given
set of input parameters and the fully converged value $E\pedex{gap} = 6.27$~eV. 
The blue line is obtained by varying one parameter at a time, while keeping the remaining
parameters to loose values. The red line corresponds
to calculations with the remaining parameters being set to stringent values.
See section~\ref{sec:conv} for a detailed discussion of all parameters 
and Table~\ref{tab:convergence} for the loose and stringent values used in the
calculation.}
\label{fig:conv}
\end{figure*}

\begin{figure*}[b]
\centering
\includegraphics{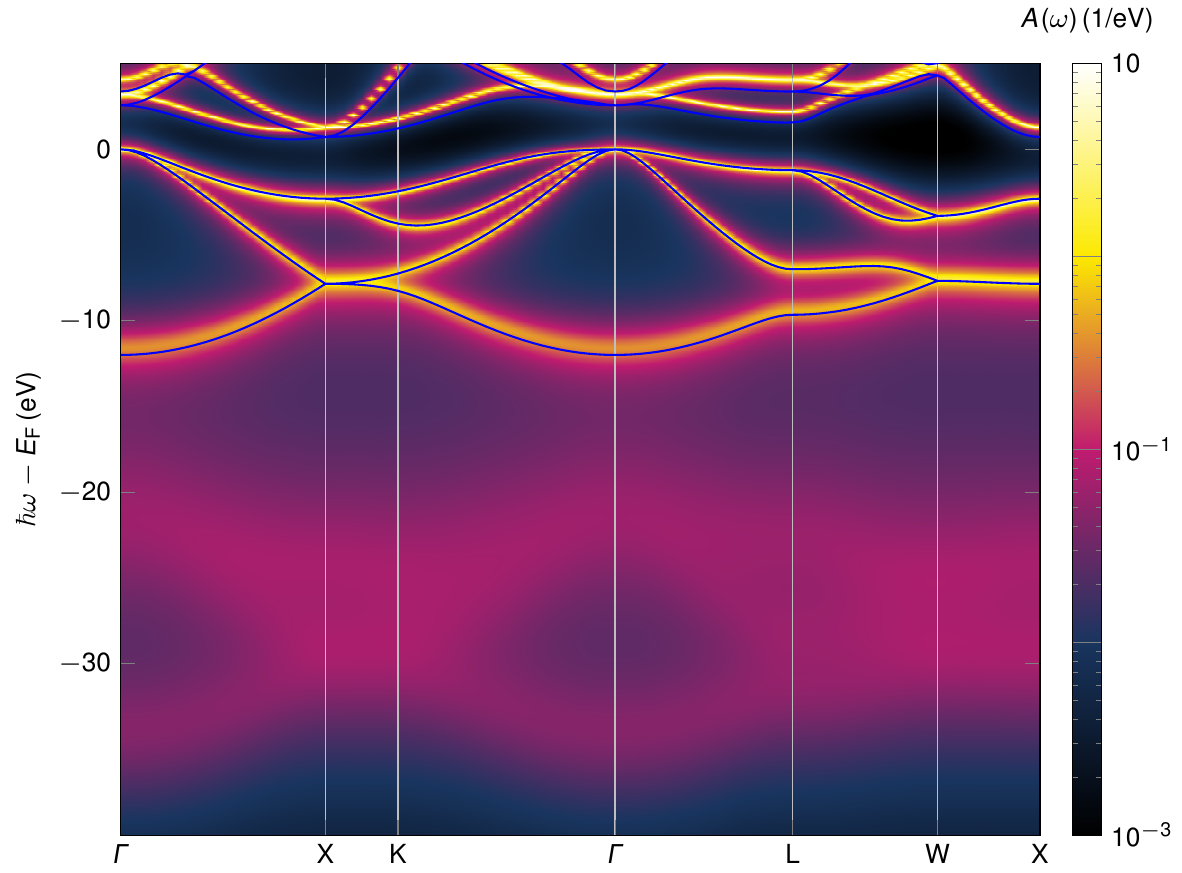}
\caption{Spectral function of silicon along high-symmetry directions in the
Brillouin zone. Results are obtained with full-frequency integration along
the imaginary axis. 
For comparison the DFT bandstructure is shown as blue
lines. Numerical parameters used for the calculations are listed in Table~\ref{tab:convergence}.}
\label{fig:spectral}
\end{figure*}

To assess the accuracy of \sgw, we verify the implementation by comparing
to reference calculations from the literature.\cite{sk07,fbsbs12,Chen2014,Chen2015b}
We use seven reference systems in the diamond (C, Si, and Ge), the zincblende (BN, CdS,
and GaAs), or rocksalt structure (MgO).
All lattice constants are taken from Ref.~\citenum{sk07}, except for 
that of Ge which is from Ref.~\citenum{Colella2009}.

The DFT ground state calculation was run with Quantum ESPRESSO v6.3 using
a $12 \times 12 \times 12$ $\vect k$-point
mesh and the PBE exchange-correlation potential.\cite{pbe96}
The pseudopotentials are verified according to the $\Delta$-test;\cite{Lejaeghere2016}
as a compromise between speed and accuracy, we employ the ONCV\cite{hama13} pseudopotentials from the 
SG15 library\cite{sg15} for compounds not including $d$ electrons, and the stringent
pseudopotentials v0.4 from \url{pseudo-dojo.org}\cite{Setten2018} otherwise.

In Table~\ref{tab:benchmark}, we present the comparison between the reference
calculations and the results obtained with \sgw~v0.15.
The \sgw{} calculations were converged using the algorithm described in
Sec.~\ref{sec:conv} resulting in the parameters listed in Table~\ref{tab:convergence}.
Overall, we find that our results agree well with the literature, within
an error bar of 0.1~eV usually quoted in the literature for $GW$ calculations.

\subsection{Converging a \emph{GW} calculation} \label{sec:conv}

Predictive $GW$ calculations require the convergence of several
numerical parameters. An
efficient strategy is required to obtain these values at a reasonable
computational cost.
In order to tackle this multi-dimensional optimization, a possible strategy
is to keep all convergence parameters at a loose setting,
and only focus on the convergence behavior of a single parameter.
To determine reasonable values for the loose setting it is a good
strategy to focus on parts of the calculation that can be evaluated separately
from the rest. First, one considers the exchange contribution to the $GW$ self-energy,
which is
the most challenging quantity to converge in \sgw{}, but at the same time its calculation
is significantly less demanding than the correlation part.
The analysis of the exchange self-energy yields estimates for
the wavefunction cutoff $E\pedex{cut}\apex{scf}$ used in the DFT calculation 
(Fig.~\ref{fig:conv}a), the exchange cutoff $E\pedex{cut}\apex{x}$ (Fig.~\ref{fig:conv}b),
and the number $N_{\vect q}$ of points in the $\vect q$-point grid used to evaluate
the Brillouin-zone integrals (Fig.~\ref{fig:conv}c).
The wavefunction and the exchange cutoff are somewhat intertwined, and occasionally
the former needs to be increased to allow for convergence of the latter.
Secondly, we investigate the convergence of the head of the dielectric 
function ($\vect q=0$ point) with respect to the number of $\vect k$ points 
$N_{\vect k}$ (Fig.~\ref{fig:conv}d)
and the coarse frequency mesh $N_{\omega,W}\apex{coarse}$ used for the analytic continuation
(Fig.~\ref{fig:conv}e). 

Guided by the initial studies of the exchange self-energy and the dielectric matrix,
we construct the loose setting for the convergence study.
In general, we will choose the smallest possible values in the smooth region of convergence.
For the correlation, we need to set three further numerical parameters:
For the dense frequency mesh $N_{\omega,W}\apex{dense}$ (Fig.~\ref{fig:conv}f), we choose
a mesh twice as dense as the coarse mesh; 
for the correlation cutoff $E\pedex{cut}\apex{c}$ (Fig.~\ref{fig:conv}g), we select
typically a third of the value used for the exchange self-energy; 
and we use just two frequencies $N_{\omega,\Sigma}\apex{coarse} = 2$ (Fig.~\ref{fig:conv}h)
to perform the analytic continuation of the correlation self-energy from the
imaginary to the real frequency axis.

With the constructed loose mesh, we then perform the convergence study
individually increasing every single one of the parameters until the quasiparticle
energies of interest are converged.
The advantage of this strategy is that the computational cost of the complete convergence
study is significantly smaller than that of the subsequent $GW$ calculation.
In Fig.~\ref{fig:conv}, we demonstrate that the estimate based on the loose
setting provides a conservative estimate of the stringent parameters 
needed to obtain converged results.

\begin{figure}
\centering
\includegraphics{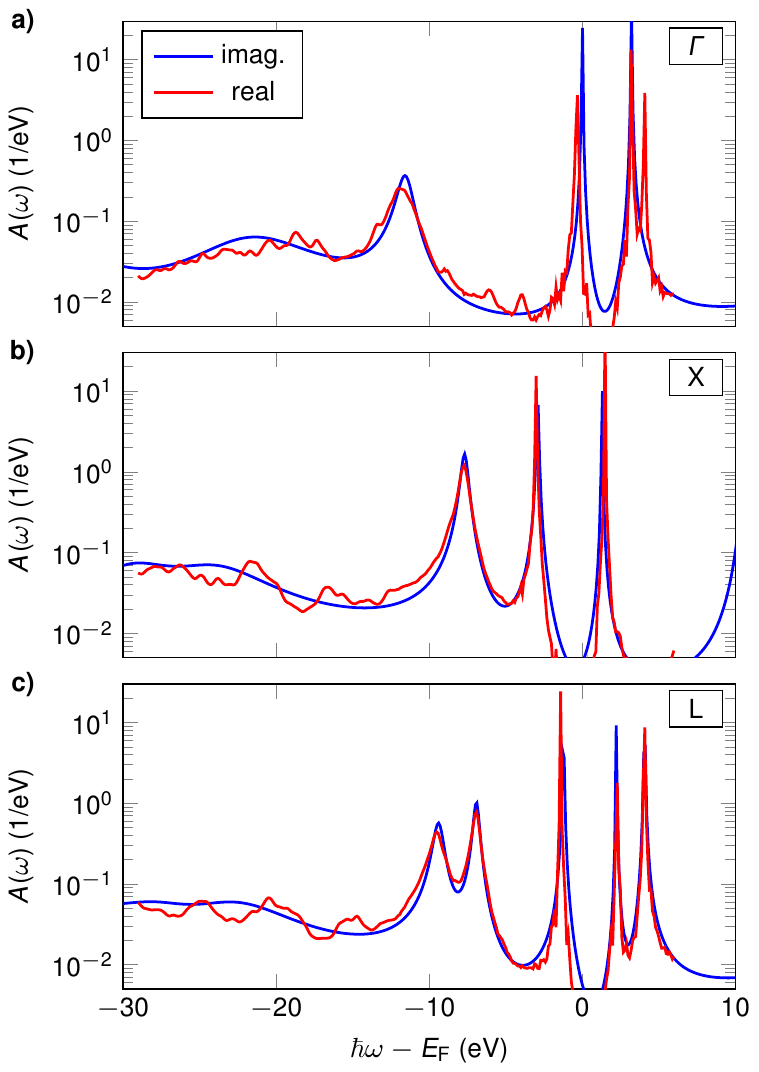}
\caption{Comparison between spectral function $A(\omega)$ of silicon at {\bf a})
the $\Gamma$ point, {\bf b}) the X point, and {\bf c}) the L point
when $\Sigma\apex{c}$ is obtained by convolution along real (red) or imaginary
(blue) frequency axis. The small shift in the valence band top is a consequence
of the finite broadening $\eta = 0.15$ eV employed in the calculation of the Green's
function. Numerical parameters used for the calculations are listed in Table~\ref{tab:convergence}.}
\label{fig:real}
\end{figure}

\subsection{Spectral function}
One noteworthy feature of \sgw{} is the possibility of calculating the $\vect k$-resolved
spectral function at any arbitrary point in the Brillouin zone.
This feature allows us to compute $GW$ band structures without the need for
interpolation techniques.
The frequency resolution provides insight into the electronic
lifetimes, and can be directly compared to ARPES experiments.
In Fig.~\ref{fig:spectral}, we show a representative calculation for
silicon.
We can see the quasiparticle bands and the weaker replica of the
bands due to the plasmon satellite.
We note that the energy and broadening of the plasmon satellites
are not accurately
described in $GW$. An accurate description of these features requires
the inclusion of higher order effects via
the cumulant expansion.\cite{Aryasetiawan1996,Holm1997,ag98,
Hedin1999,Gumhalter2005,Gumhalter2012,Zhou2015,Verdi2017,Caruso2018}

\subsection{Real frequency integration}

\begin{figure}
\centering
\includegraphics{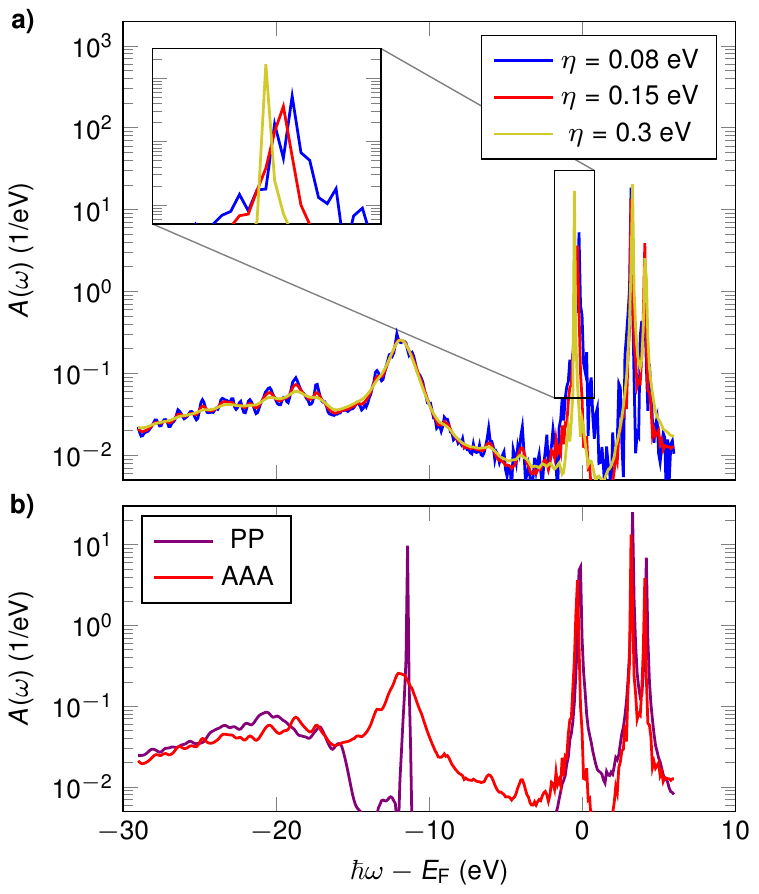}
\caption{Impact of numerical technique on spectral function of silicon
at $\Gamma$ obtained by
convolution along the real frequency axis. 
{\bf a)} Increasing the smearing from $\eta = 0.08$~eV (blue) to 0.15~eV (red)
or 0.3~eV (yellow) results in a decrease of the noise but also a shift of the
valence band maximum to lower energies (see inset). 
{\bf b)} Using the plasmon pole (PP) model (purple) instead of the AAA analytic 
continuation (blue) results in vanishing linewidths. We note that
the vertical axes are on a logarithmic scale.
Numerical parameters used for the calculations are listed in 
Table~\ref{tab:convergence}.}
\label{fig:technical}
\end{figure}

As outlined in Sec.~\ref{sec:corr}, \sgw{} provides the option to perform
the frequency convolution along the real or the imaginary frequency axis.
In Fig.~\ref{fig:real}, we demonstrate that these two implementations give
nearly identical results for the spectral function at high-symmetry points of
silicon.
The main identifiable differences is a small shift of the valence band top
to lower energies and an increased noise when using integration along the
real axis.
Both of these differences originate from the small parameter $\eta = 0.15$~eV that
is used to shift the frequency slightly off the real axis. This shift is
necessary in order to avoid the singularities on the real axis, so as to obtain numerically stable results.
On the one hand, if $\eta$ is decreased to 0.08~eV the resulting spectral function
becomes very noisy (see Fig.~\ref{fig:technical}a).
On the other hand, if $\eta$ is increased to 0.3~eV, the energy of the  
quasiparticle becomes inaccurate (see inset of Fig.~\ref{fig:technical}a).

In general, the integration along the imaginary axis with subsequent analytic
continuation is one to two orders of magnitude cheaper because the Green's 
function and the screened Coulomb interaction have less structure there.
As the results of both methods are very similar, we chose the
integration along the imaginary axis as the default in \sgw{}, and we 
provide the integration along the real axis as an optional feature
for validation.

\begin{figure}
\centering
\includegraphics{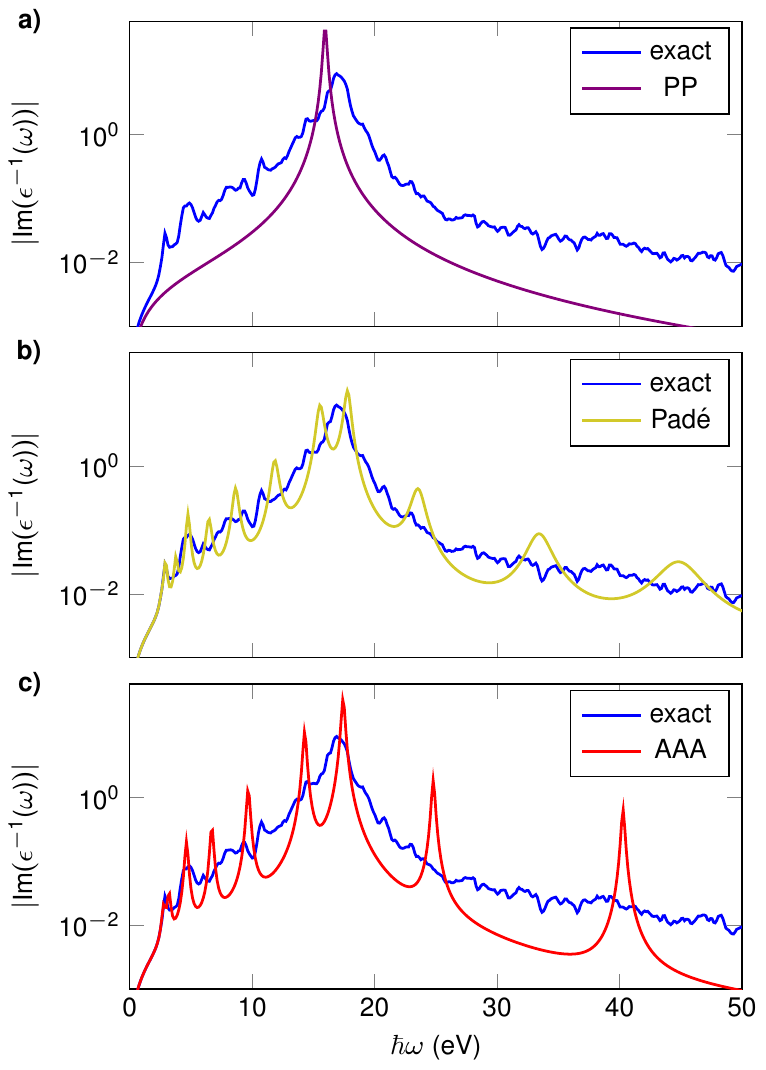}
\caption{Inverse of the dielectric matrix of silicon at the $\Gamma$ point obtained
by solving the Sternheimer equation~\eqref{eq:wavef} along the real axis (blue) compared to
{\bf a)} plasmon pole model (purple), {\bf b)} Pad\'{e} expansion\cite{vs77} (yellow),
and {\bf c)} AAA algorithm (red).\cite{Nakatsukasa2018}
For the analytic continuation with either Pad\'{e} or AAA,
201 frequencies along the imaginary axis were evaluated.}
\label{fig:analytic}
\end{figure}

In Fig.~\ref{fig:technical}b, we show that obtaining accurate quasi-particle
lifetimes requires going beyond the plasmon-pole approximation,
because in this approximation there is only one sharp pole at 
the plasmon frequency (cf.~Fig.\ref{fig:analytic}a).
In contrast, analytic continuation techniques can produce multiple poles and
thereby capture the broadening of the quasiparticle spectra.
In Fig.~\ref{fig:analytic}b, we show that the Pad\'e expansion\cite{vs77} can
reproduce the shape of the dielectric function more accurately 
than the plasmon-pole approximation, since it naturally incorporates multiple poles.
However,
this Pad\'{e} algorithm is numerically not stable when using a large number of 
frequencies, therefore it
cannot be used for the convolution along the real frequency axis.
To address this issue, we employ the AAA analytic\cite{Nakatsukasa2018} continuation
technique, whereby only significant frequencies are selected via a singular value 
decomposition.
Subsequently, poles with weak residues are removed so that only the most
important contributions remain.
In Fig.~\ref{fig:analytic}c, we show that the AAA method reduces the number of poles
compared to the Pad\'{e} expansion while still reproducing the overall shape and 
magnitude of the dielectric matrix.

\section{Conclusion}
We presented the \sgw{} code, which calculates
many-body $GW$ self-energies, quasiparticle band structures, and spectral
functions
by solving linear-response Sternheimer equations.
The linear response scheme allows us to calculate the $GW$ self-energy at
arbitrary $\vect k$ points. This is
particularly useful for computing band structures without resorting
to interpolation techniques, and for comparison to ARPES
experiments. \sgw{} goes beyond the plasmon-pole approximation
by performing full-frequency integration along the real or the imaginary axis.

\sgw{} is aligned with the version 6.3 of Quantum Espresso. The
dedicated website \url{http://www.sternheimergw.org} provides users
with installation guidelines, tutorials, and a comprehensive documentation
of all input variables.
The code is efficiently parallelized and has been tested extensively
on high-performance computing architectures.
Updated versions of the code are distributed in the GitHub repository
\url{https://github.com/QEF/SternheimerGW}.

\section*{Acknowledgments}
The development of \sgw{} has received funding from 
the Leverhulme Trust (Grant RL-2012-001), the
Graphene Flagship (Horizon 2020 Grant No. 785219 - GrapheneCore2), and the
UK Engineering and Physical Sciences Research Council (Grant No. EP/M020517/1).
Furthermore, the authors acknowledge
the use of the University of Oxford Advanced Research Computing (ARC) 
facility (http://dx.doi.org/10.5281/zenodo.22558),
the ARCHER UK National Supercomputing Service under the “T-Dops” project,
the Cambridge Service for Data Driven Discovery (CSD3) funded by EPSRC Tier-2 (Grant EP/P020259/1),
the DECI resource “Cartesius” based in the Netherlands at SURFsara and “Abel”
based in Oslo with support from the PRACE AISBL,
and PRACE for awarding us access to MareNostrum at BSC-CNS, Spain.



%
%


\bibliographystyle{elsarticle-num}
\bibliography{biblio.bib}







\end{document}